\documentclass[12pt,aps,amsmath,amssymb,prb,reprint,showkeys] {revtex4-2}
\usepackage{graphicx, textcomp, wasysym, amssymb, epsfig}

\begin{document}
\title{Complex magnetic phase diagram of metamagnetic MnPtSi}

\author{M. B. Gam\.za$^{1,2}$ \email{MGamza@uclan.ac.uk}, W. Schnelle$^{2}$, H. Rosner$^{2}$, S.-V.~Ackerbauer$^{2}$, Yu. Grin$^{2}$, and A. Leithe-Jasper$^{2}$}

\affiliation{$^{1}$Jeremiah Horrocks Institute for Mathematics, Physics and Astrophysics, University of Central Lancashire, Preston PR1 2HE, UK}
\affiliation{$^{2}$Max--Planck Institute for Chemical Physics of Solids, 01187 Dresden, Germany}


\begin{abstract}

The magnetic, thermal and transport properties as well as electronic band structure of MnPtSi are reported. MnPtSi is a metal that undergoes a ferromagnetic transition at $T_{\mathrm{C}}=340$(1)~K and a spin--reorientation transition at $T_{\mathrm{N}}=326$(1)~K to an antiferromagnetic phase. First--principles electronic structure calculations indicate a not--fully polarized spin state of Mn in a $d^5$ electron configuration with $J=S=3$/2, in agreement with the saturation magnetization of \mbox{3~$\mu_{\mathrm{B}}$} in the ordered state and the observed paramagnetic effective moment. A sizeable anomalous Hall effect in the antiferromagnetic phase alongside the computational study suggests that the antiferromagnetic structure is non--collinear. Based on thermodynamic and resistivity data we construct a magnetic phase diagram. Magnetization curves $M$($H$) at low temperatures reveal a metamagnetic transition of spin--flop type. The spin--flopped phase terminates at a critical point with $T_{\mathrm{cr}}\approx 300$~K and $H_{\mathrm{cr}}\approx 10$~kOe, near which a peak of the magnetocaloric entropy change is observed. Using Arrott plot analysis and magnetoresistivity data we argue that the metamagnetic transition is of a first--order type, whereas the strong field dependence of $T_{\mathrm{N}}$ and the linear relationship of the $T_{\mathrm{N}}$ with $M^2$ hint at its magnetoelastic nature.

\end{abstract}


\maketitle

\section{Introduction}

The discovery of a rare Lifshitz multicritical behaviour in MnP \cite{Lifszyc, MnPlast} initiated an extensive study on the family of related Mn--based ternary compounds adopting orthorhombic crystal structures of the TiNiSi--type.\cite{MnIrSi, MnCoGe, MnNiGe, MnNiSi, MnRhSi, Barcza, MnCoSipierw, MnCoGestr, Gercsi, structuralne, FeMnPSi, Morrison, magnetocaloricMnCoSi, Morrison2, magnetocaloriccoupled, Kun, Barcza2} Magnetic structures of these materials are ranging from a commensurate antiferromagnetic (AFM) ordering through collinear and non--collinear incommensurate helical, cycloidal and fan spin structures to simple ferromagnetic (FM) states.\cite{MnCoGe, MnNiSi, MnNiGe, MnRhSi, MnIrSi, MnCoSipierw} The appearance of non--collinear spin arrangements makes these compounds of interest because of their potential for future applications in spintronics.\cite{X}

Numerous studies have shown that magnetism of the Mn--based TiNiSi--type compounds can be tuned by changing distances between neighbouring Mn species.\cite{Gercsi, Barcza, FeMnPSi, Barcza2} Here, the strong connection between crystal structures and magnetic properties is thought to arise, unusually, from competing interatomic exchange interactions.\cite{Barcza, FeMnPSi} The magnetoelastic coupling can cause an Invar--like effect in sample volume.\cite{Kun, Barcza} Furthermore, it can also bring about first--order magnetoelastic phase transitions that release a large entropy over a narrow temperature range, even though the symmetry of the crystal lattice is the same on both sides of the phase transition.  
In MnCoSi, for instance, the thermal evolution of the helical AFM state is accompanied by a huge and opposing change in the two shortest Mn--Mn distances of $\sim$2\% that not only gives rise to an Invar--like behaviour but, in finite magnetic fields, it couples to the suppression of the helimagnetism and is believed to be the precursor to a metamagnetic tricritical point with strongly enhanced magnetostrictive and inverse magnetocaloric effects.\cite{Barcza, Morrison2}

The critical behaviour near first--order magnetoelastic transitions can be controlled by changing chemical composition or annealing conditions.\cite{Morrison, Morrison2, magnetocaloricMnCoSi, magnetocaloriccoupled, magnetocaloricMnP} Importantly, the thermal hysteresis can be often tuned to reach small values while maintaining a large magnetocaloric effect.\cite{Morrison, magnetocaloricMnCoSi} Consequently, first--order magnetoelastic transitions provide a promising venue for producing magnetocaloric materials that could be used as magnetic refrigerants operating at high thermal cycling frequencies.\cite{Trung}

Searching for new members of the TiNiSi--type family with first--order magnetoelastic transitions and/or intriguing magnetic structures, we synthesized MnPtSi.\cite{PNAS-Rosner} Manganese in MnPtSi shows magnetic moment of 3~$\mu_{\mathrm{B}}$. Remarkably, we found that this local magnetic polarization prevents the formation of Mn--Mn bonds and thus rules the adopted TiNiSi--type crystal structure. Here, we present the results of thermodynamic and transport measurements on polycrystalline MnPtSi aiming at exploring a magnetic phase diagram. 
Our study shows that MnPtSi undergoes a FM transition at $T_{\mathrm{C}}$=~340(1)~K and a spin--reorientation transition at $T_{\mathrm{N}}$=~326(1)~K to an AFM phase.  
A sizeable anomalous Hall effect (AHE) observed in the AFM state suggests that the low--$T$ magnetic ordering is non--collinear. 
The experimental data is supplemented by first--principles electronic structure calculations which, in conjunction with magnetization data, are used to address the spin state of Mn. Finally, we discuss the nature of the magnetic phase transitions and the role of magnetoelastic interactions in MnPtSi.

\section{\label{methods}Methods}

Polycrystalline samples of MnPtSi were prepared as described in \onlinecite{PNAS-Rosner}. The specimens were examined by means of powder x-ray diffraction measurements and metallographic study, and were found to be single--phase.

\begin{figure}
\centering
\includegraphics[width=0.47\textwidth,angle=0]{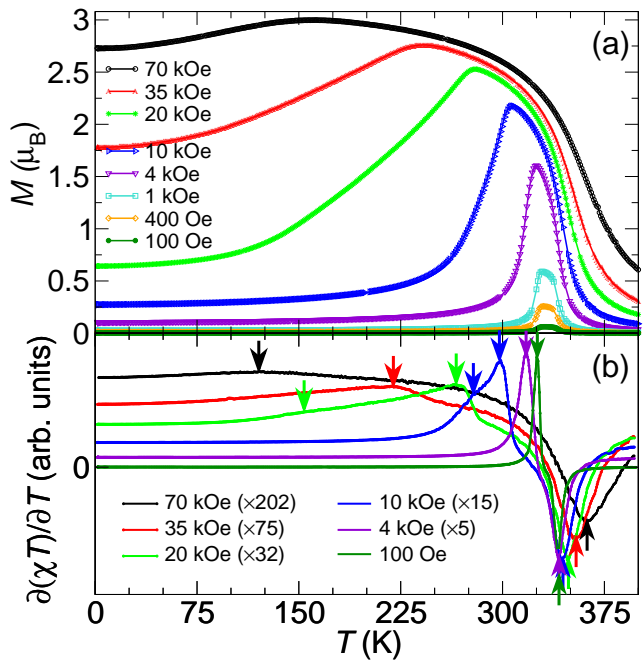}
\caption{\label{fig:Fig1} (Color online) (a) Magnetization per f.u. versus temperature in selected magnetic fields. (b) $\partial$($\chi \cdot T$)/$\partial T$ vs $T$ estimated from the $M$($T$) data and scaled to allow visual comparisons of curves for different fields. Arrows indicate magnetic transitions.}
\end{figure} 

Dc magnetization measurements were carried out in a SQUID magnetometer (MPMS~XL--7, Quantum Design) in the temperature range of 1.8--750~K. In addition, pulsed--field magnetization measurements were performed in the High Magnetic Field Laboratory at Helmholtz Zentrum Dresden--Rossendorf in magnetic fields up to 60~T. Details of the experimental procedure are described in \onlinecite{Rosendorf}. Heat capacity was determined by a relaxation--type method (HC option, PPMS-9, Quantum Design). ACT option of the PPMS-9 was used for electrical resistivity and Hall effect measurements on polycrystalline blocks. The Hall resistivity data collected during increasing and subsequent decreasing field between -9~T and +9~T did not display any magnetic hysteresis effect. The conventional antisymmetrization method was used to correct for small symmetric signals superimposed on the Hall resistivity. 

First principles electronic band structure calculations were performed using the full--potential local--orbital code FPLO (version 9.01--35) \cite{FPLO} using experimental lattice and atomic positional parameters obtained from room--temperature x--ray diffraction studies.\cite{PNAS-Rosner} In the fully relativistic calculations the four--component Kohn--Sham--Dirac equation containing implicitly spin--orbit coupling up to all orders was solved self--consistently. Within the local (spin) density approximation of the density functional theory, the exchange--correlation potential in the form proposed by Perdew and Wang\cite{LSDAPW} was employed. A dense $k$--mesh (15$\times$20$\times$13, 616~points in the irreducible wedge of the Brillouin zone) was used to ensure accurate density of states (DOS) and total energy information. Several collinear AFM or ferrimagnetic (FIM) spin configurations were considered based on supercells containing up to eight formula units of MnPtSi with up to four crystallographically nonequivalent atomic positions occupied by Mn.      
In these calculations similar densities for $k$--mesh were assumed. As a first approximation to simulate a paramagnetic (PM) state, the disordered local moments (DLM)\cite{DLM} approach was used, in which thermal disorder among magnetic moments is described using the coherent potential approximation (CPA)\cite{CPA}. The scalar--relativistic CPA calculations were performed assuming that the Mn site was occupied randomly by equal numbers of Mn atoms with opposite spin--polarization directions.

\section{Results}

\subsection{Magnetic measurements}
\label{Magnetization}

\begin{figure}
\centering
\includegraphics[width=0.47\textwidth,angle=0]{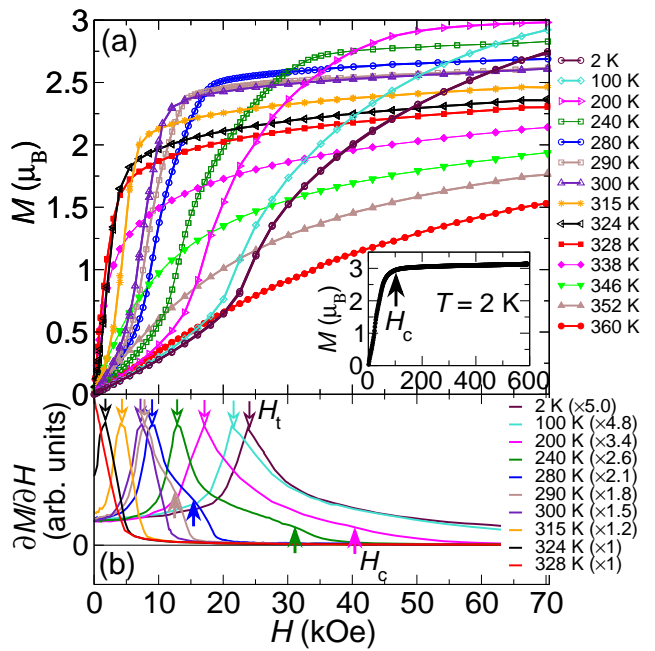}
\caption{\label{fig:Fig2} (Color online) (a) Isothermal magnetization curves (per f.u.) measured at selected temperatures. The inset shows $M$($H$) data recorded at 2~K in pulsed magnetic fields up to 600~kOe. (b) Field dependencies of \mbox{$\partial M$/$\partial H$}. Arrows indicate positions of peaks in the \mbox{$\partial M$/$\partial H$} vs $H$ curves that are used as estimates for $H_{\mathrm{t}}$ and $H_{\mathrm{c}}$ at different temperatures. } 
\end{figure} 

\begin{figure*}
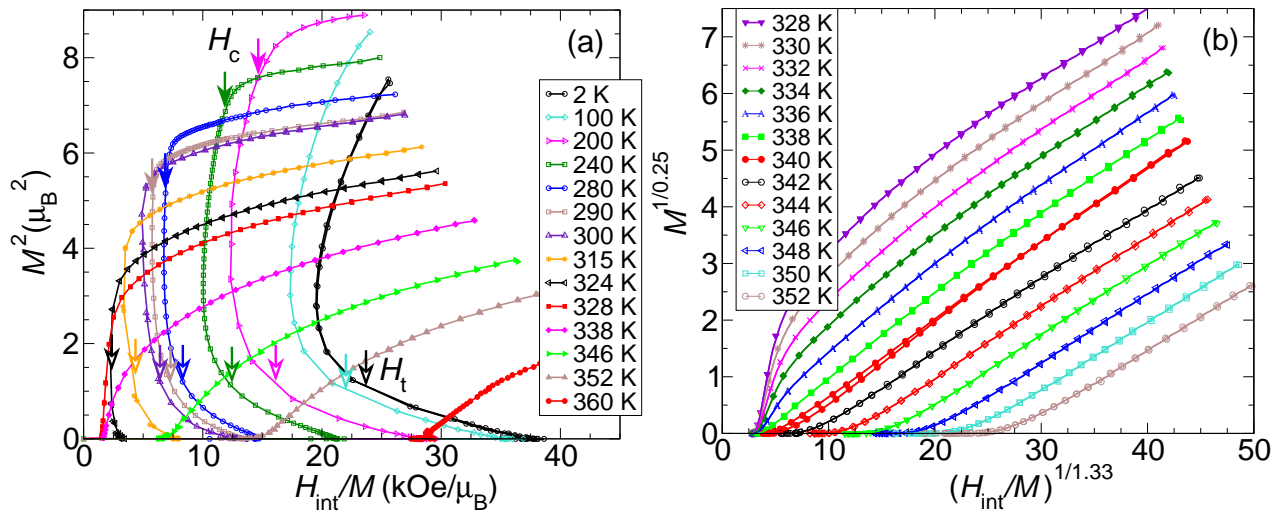

\includegraphics[width=0.47\textwidth,angle=0]{Fig3p1.eps}
\includegraphics[width=0.46\textwidth,angle=0]{Fig3p2.eps}
\caption{\label{fig:Fig3} (Color online) (a) Standard Arrott plot with corrected $M$($H$) curves (per f.u.) measured at selected temperatures. Points corresponding to magnetic fields $H_{\mathrm{t}}$ and $H_{\mathrm{c}}$ at different temperatures are indicated by open and filled arrows, respectively. (b) Modified Arrott plot with critical exponents $\beta=0$.25 and $\gamma=1$.33 applied to get straight lines in the high field range. }
\end{figure*}

\begin{figure}
\includegraphics[width=0.45\textwidth,angle=0]{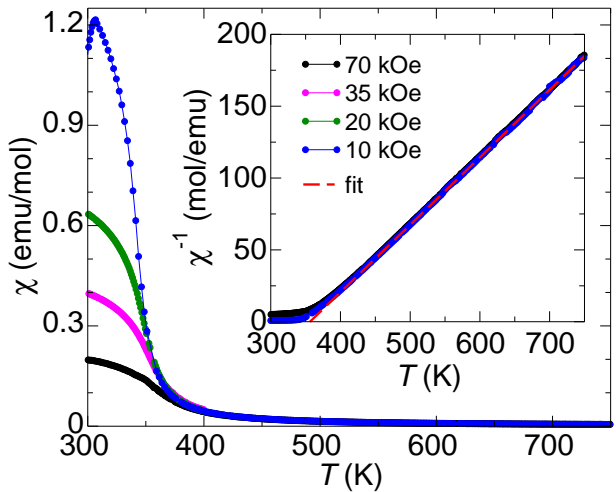}
\caption{\label{fig:Fig4} (Color online) High--temperature magnetic susceptibility versus temperature measured in selected magnetic fields. Inset shows $\chi^{-1}$($T$), together with the fit using Eq.~\ref{eq:Equ1}.}
\end{figure}

Figures~\ref{fig:Fig1},~\ref{fig:Fig2}, ~\ref{fig:Fig3}, and~\ref{fig:Fig4} present results of a magnetization study on polycrystalline MnPtSi. The temperature dependence of the magnetization $M$($T$) (Fig.~\ref{fig:Fig1}) indicates that MnPtSi undergoes two successive magnetic phase transitions. A rapid increase in $M$($T$) at $T_{\mathrm{C}}\approx 340$~K is indicative of an onset of a FM--type order. In turn, a sudden decrease in the magnetization that in weak magnetic fields has the steepest slope at $T_{\mathrm{N}}\approx 326$~K points to a spin--reorientation transition. The latter effect is gradually suppressed by the applied magnetic field, whereas the FM transition moves slowly toward higher $T$ with increasing $H$.

To evaluate the effect of magnetic field on the magnetic transitions in more detail, magnetic specific heat was estimated from the $M$($T$) data using Fisher's relation\cite{Fisher}, \mbox{$C_{\mathrm{m}}$~$\propto$~$\partial$($\chi \cdot T$)/$\partial T$}, where $\chi=M$/$H$.
The resulting $C_{\mathrm{m}}$($T$) curves are plotted in Fig.~\ref{fig:Fig1}b. They show a pronounced peak due to the AFM transition and a deep minimum marking the onset of the FM order. Both these effects broaden in magnetic fields, and the AFM peak moves quickly toward low $T$, whereas the FM feature shifts slowly to higher $T$. Interestingly, the AFM peak splits into two distinct components that are visible in the $C_{\mathrm{m}}$($T$) curves for fields of 10--20~kOe, suggesting that an additional magnetic phase transition sets in.

To further explore the complex magnetic behaviour, isothermal magnetization was measured at a number of temperatures. Selected results are shown in Fig.~\ref{fig:Fig2}a. The $M$($H$) data collected during increasing and subsequent decreasing field do not display any magnetic hysteresis effect in the entire investigated temperature range. At high temperatures the shape of the $M$($H$) curves evolves from straight lines expected for a PM state to strongly bend curves indicative of a FM ordering at temperatures slightly above 326~K. Below $T_{\mathrm{N}}\approx 326$~K, in low fields the isothermal magnetization changes very slowly with $H$ and there is no remanence at $H=0$, as expected for an AFM ordering. Importantly, the $M$($H$) curves show a rapid increase starting at finite magnetic fields, implying a metamagnetic transition. The threshold magnetic field denoting the AFM phase boundary, $H_{\mathrm{t}}$, is of 24~kOe for $T$~=~2~K (Fig.~\ref{fig:Fig2}a). It decreases with increasing $T$, and finally the metamagnetic transition ceases at $T_{\mathrm{N}}$.

We note that at low $T$ the raise in the isothermal magnetization near $H_{\mathrm{t}}$ is only of 0.6~$\mu_{\mathrm{B}}$ and the $M$($H$) saturates at fields much higher than the threshold field $H_{\mathrm{t}}$ (Fig.~\ref{fig:Fig2}a). Importantly, the shape of the low temperature $M$($H$) curves resembles those expected for spin--flop antiferromagnets. 
The critical field at which the material becomes fully magnetized, $H_{\mathrm{c}}$, is of 100~kOe for $T$~=~2~K (Fig.~\ref{fig:Fig2}a, inset), which is over four times larger that the $H_{\mathrm{t}}$, but it decreases quickly with increasing $T$. Remarkably, at $T_{\mathrm{cr}}\approx300$~K the spin--flopped (SF) phase terminates and the metamagnetic transition turns into spin--flip type. This alteration of the metamagnetic behaviour is evidenced by the change in shape of the $\partial M$/$\partial H$ curves from a double--peak structure to a single peak form (Fig.~\ref{fig:Fig2}b).

To get an insight into the nature of the magnetic transitions in MnPtSi, the Arrott plot analysis of the $M$($H$) curves was performed. To this end, the $M$($H$) data was corrected for the demagnetization effect\cite{note} and redrawn as $M^2$ versus $H/M$ (Fig.~\ref{fig:Fig3}a). According to Banerjee\cite{Banerjee}, the slope of lines in the Arrott plot indicates the order of the phase transition: negative curvatures correspond to a first--order transition, positive to a second--order one. Thus, negative slopes of the standard Arrott curves for MnPtSi at the points corresponding to the fields $H_{\mathrm{t}}$ indicate that the AFM phase boundary is of a first--order type. In turn, positive curvatures of the Arrott curves for $T < T_{\mathrm{cr}}$ in the high field range suggest that the SF--FM transition is second--order in nature. Further, for $T > T_{\mathrm{N}}$ the slopes of the entire $M^2$($H/M$) curves are positive, implying a second--order character of the FM--PM transition in MnPtSi.

The standard Arrott plot analysis does not allow to determine the FM--PM transition temperature because all curves in Fig.~\ref{fig:Fig3}a for $T\approx340$~K show pronounced curvature even at high magnetic fields. 
Therefore, we applied the modified Arrott plot technique\cite{Arrott}. Accordingly, the corrected $M$($H$) data was plotted as $M^{1/\beta}$ versus ($H$/$M$)$^{1/\gamma}$ using various $\beta$ and $\gamma$ values until straight lines parallel to each other were obtained at high magnetic fields (Fig.~\ref{fig:Fig3}b). Since in the low field range the isotherms are curved downwards for $T > 340$~K and upwards for $T < 338$~K, we conclude that $T_{\mathrm{C}}\approx 339$(1)~K. However, the estimated critical exponents $\beta \approx 0.25$ and $\gamma \approx 1.33$ are very different from values predicted by the mean--field theory as well as 3D~Heisenberg or Ising models.\cite{exponents} Importantly, the lines in Fig.~\ref{fig:Fig3}b do not pass through the origin of the plot. Instead, they converge at $H_{\mathrm{int}}$/$M$~$\gg$~0. Such a behaviour was observed for FM materials with a significant magnetocrystalline anisotropy\cite{Landau} and for systems in which the FM transition is broadly spread both in temperature and in magnetic field due to competing FM and AFM interactions.\cite{Arrottw2} Further studies including direction--dependent thermodynamic measurements on crystals are needed to investigate critical behaviour near the magnetic transitions and to inspect the role of magnetocrystalline anisotropy in MnPtSi.

For $T > 375$~K the magnetic susceptibility does not depend on the applied magnetic field (Fig.~\ref{fig:Fig4}) and can be well described by a modified Curie-Weiss law:
\begin{equation}
\chi = \chi_0 + \frac{C}{T - \theta_{\mathrm{P}}} \:.
\label{eq:Equ1}
\end{equation}
The least--squares fit to the data in the temperature range of 450--750~K yields: $\chi_{\mathrm{0}}\approx 1\times 10^{-5}$~emu/mol, \mbox{$\theta_{\mathrm{P}}\approx 352$~K,} and $C\approx 2.155$~emu~K/mol that corresponds to the effective paramagnetic moment $p_{\mathrm{eff}}\approx$~4.15~$\mu_{\mathrm{B}}$ per f.u.
The estimated paramagnetic Curie temperature of $\sim$352~K is slightly higher than the $T_{\mathrm{C}}\approx340$~K and thus confirms the dominance of FM interactions between magnetic moments of Mn in MnPtSi. The positive $\chi_{\mathrm{0}}$ hints at metallic properties, in agreement with electrical resistivity measurements (Section \ref{Resistivity}). Importantly, the effective moment and the saturation magnetization in the ordered state of slightly over 3~$\mu_{\mathrm{B}}$ per f.u. (Fig.~\ref{fig:Fig2}a, inset) unanimously indicate that Mn in MnPtSi adopts a $J=3$/2 configuration, in agreement with \onlinecite{PNAS-Rosner}.

\subsection{Specific heat}
\label{SpecificHeat}

\begin{figure}
\includegraphics[width=0.47\textwidth,angle=0]{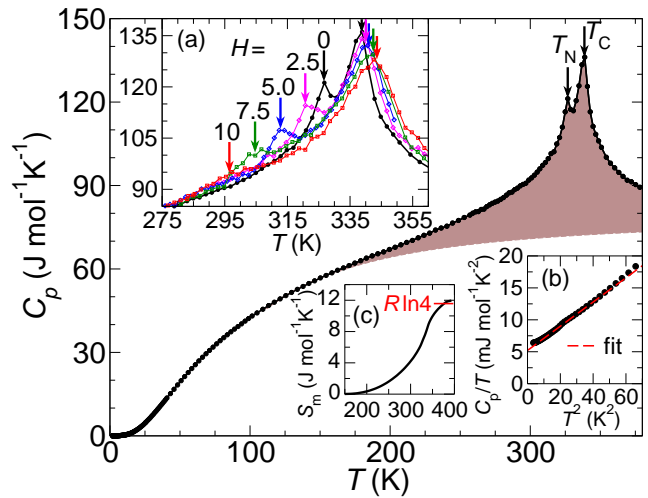}
\caption{\label{fig:Fig5} (Color online) Specific heat of MnPtSi. Magnetic phase transitions are indicated. The brown area represents the magnetic specific heat estimated as described in the text. Insets: (a) Specific heat in selected external magnetic fields. The values of $H$ (kOe) are given above the arrows indicating the transition temperatures; (b) $C_{\mathrm{p}}$/$T$ plotted versus $T^2$. The red dashed line represents the fit to the experimental data using Eq.~\ref{eq:Equ2}; (c) Magnetic entropy.}
\end{figure}

Fig.~\ref{fig:Fig5} presents the specific heat of MnPtSi. There are two distinct $\lambda$--type anomalies in $C_p$($T$) implying that MnPtSi undergoes two successive phase transitions. The higher peak at $T_{\mathrm{C}}\approx 339$~K in zero field concurs with the steep increase in the $M$($T$) and thus can be attributed to the FM transition. As shown in Fig.~\ref{fig:Fig5}a, increasing magnetic fields shift this anomaly towards higher temperatures, as expected based on the $M$($T$) curves (Fig.~\ref{fig:Fig1}). 
The second anomaly in $C_p$($T$) at $T_{\mathrm{N}}\approx 326$~K in $H=0$ coincides with the rapid decrease in the $M$($T$) observed in weak magnetic fields and is ascribed to the spin--reorientation transition. This transition is very sensitive to the applied magnetic field: it moves towards lower $T$ and broadens strongly with increasing $H$, in agreement with the magnetization data (Section~\ref{Magnetization}).

To evaluate the electronic specific heat, we replotted $C_p$($T$) for MnPtSi as $C_p$/$T$ vs $T^2$ (Fig.~\ref{fig:Fig5}b). For $T<7$~K the experimental data is well described by:
\begin{equation}
C_p/T=\gamma +\beta T^2   \:
\label{eq:Equ2}
\end{equation}
with the Sommerfeld coefficient $\gamma$~$\approx$~5.3~mJ~mol$^{-1}$K$^{-2}$  and $\beta$~=~0.189~mJ~mol$^{-1}$K$^{-4}$. The latter value corresponds to an initial Debye temperature of 314~K on the presumption that magnetic excitations do not contribute to the specific heat at such low temperatures and therefore the second term in Eq.~\ref{eq:Equ2} describes the Debye $T^3$ approximation of the lattice specific heat. This premise is justified by small and nearly temperature independent low--field magnetization in this $T$ range (Section~\ref{Magnetization}) indicating that magnetic excitations are strongly suppressed in the AFM phase at low temperatures.

To inspect the magnetic part of the specific heat, first we need to evaluate the lattice contribution $C_{\mathrm{l}}$($T$).
Unfortunately, our efforts to find a nonmagnetic compound suitable as a phonon reference were unsuccessful. For the series of $M$Pt\{Si,Ge\} ($M$=Ti,V,Cr,Zr) adopting the TiNiSi--type structure distinct changes in phonon DOSs are observed, which are the subject of a separate study.\cite{GPH} Attempts to fit the $C_p$($T$) of MnPtSi as the sum of the electronic part given by $\gamma T$ and the $C_{\mathrm{l}}$($T$) approximated by the Debye model did not give satisfactory results (not shown). Since the magnetization data (Section~\ref{Magnetization}) indicates that pronounced magnetic excitations develop in the AFM phase only for $T>100$~K, a rough estimate of the lattice specific heat was made assuming that \mbox{$C_{\mathrm{l}}$($T$) = $C_p$($T$) - $\gamma T$} for $T<100$~K, while at higher temperatures the $C_{\mathrm{l}}$($T$) is described using the Debye model with \mbox{$\theta_{\mathrm{D}} \approx 375$~K}. Subtracting the obtained $C_{\mathrm{l}}$($T$) and the $\gamma T$ contributions from the total $C_p$($T$) gives the magnetic specific heat depicted  as a brown area in Fig.~\ref{fig:Fig5}. The resulting magnetic entropy $S_{\mathrm{m}}$($T$) is plotted in Fig.~\ref{fig:Fig5}c. The $S_{\mathrm{m}}$($T$) saturates in the PM state at $\approx$12~J/(mol~K), which is close to the entropy of $R$ln4 anticipated for a material with local magnetic moments of $J=3$/2. Consequently, the specific heat study provides further support for the $J=3$/2 state of Mn in MnPtSi indicated by the magnetization measurements.

\subsection{Electrical resistivity}
\label{Resistivity}

\begin{figure}
\includegraphics[width=0.47\textwidth,angle=0]{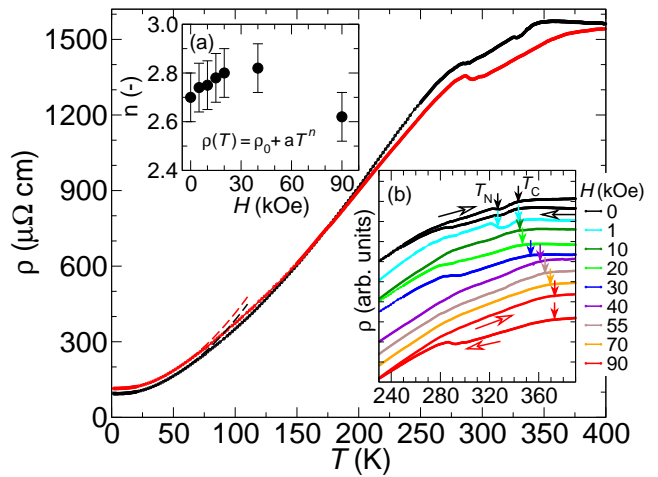}
\caption{\label{fig:Fig6} (Color online) Resistivity of MnPtSi measured on cooling in $H = 0$ (black dots) and in $H = 90$~kOe (red dots), with fits to $\rho(T)=\rho_{\mathrm{0}}+AT^2$ in the temperature range of 25--60~K (dashed lines). Insets: (a) Exponents obtained from fits of $\rho$($T$) in different magnetic fields for $T<20$~K using a power law; 
(b) $\rho$($T$) near the magnetic transitions measured in $H = 0$ and in selected fields. The curves were shifted along the vertical axis to enable their comparison. Transition temperatures $T_{\mathrm{N}}$ and $T_{\mathrm{C}}$ are indicated by filled arrows. Open arrows show the directions of measurements.}
\end{figure}

Electrical resistivity of MnPtSi shows a metallic behaviour (Fig.~\ref{fig:Fig6}). Although the absolute values of the resistivity are rather large, the residual resistivity ratio $RRR=\rho$(400~K)/$\rho$(2~K) $=15$ implies high quality of the polycrystalline specimen. We note that in silicides the resistivity is often observed to vary significantly due to the presence of insulating phases, $e.g.$ SiO$_2$, at the grain boundaries.\cite{clathrates}

There are distinct features in the $\rho(T)$ curves at temperatures close to $T_{\mathrm{C}}\approx 339$~K and $T_{\mathrm{N}}\approx 326$~K. Near the PM to FM transition the resistivity drops by about 5\%, presumably due to the loss of spin--disorder scattering. In contrast, the FM to AFM transition is associated with an increase in $\rho$($T$) that hints at an opening of a small gap in some of the electronic bands at the Fermi energy ($E_{\mathrm{F}}$). The effect of applied magnetic field on the magnetic anomalies in $\rho(T)$ illustrated in Fig.~\ref{fig:Fig6}b is consistent with results of the thermodynamic study presented in Sections \ref{Magnetization} and \ref{SpecificHeat}.

Below $\sim$20~K the temperature dependencies of resistivity measured with and without applied magnetic field can be well described by a power law with the exponent $n=2.7$(2) that only slightly changes with the field, as shown in Fig.~\ref{fig:Fig6}a. The obtained $n$ is notably smaller than $n=5$ anticipated for conventional electron--phonon (e--ph) scattering\cite{BlochG} or for scattering of electrons on AFM spin--waves\cite{AFM-spin-waves}, but it is larger than $n=2$ expected when electron--electron (e--e) scattering is dominant\cite{FL}. The exponent close to 3 suggests that at low temperatures $\rho(T)$ is governed by e--ph scattering involving $s$--$d$ transitions\cite{Wilson}, thus indicating the presence of narrow $d$--electron bands near $E_{\mathrm{F}}$.

For 25~K~$<T<60$~K the resistivity curves follow a $T^2$ behaviour, regardless of the applied magnetic field. Since in this temperature range MnPtSi adopts two distinct magnetic phases (AFM for $H<23$~kOe and SF for $H>23$~kOe) that should have different magnon dispersion relations, we conclude that scattering of electrons on spin--waves does not have a significant influence on $\rho$($T$). The observed $T^2$ dependence hints at the dominance of a Baber--type e--e scattering characteristic for Fermi liquids\cite{FL}. The coefficient of the $T^2$ term $A\approx0.031$(2)~$\mu\Omega$~cm/K$^2$ is about three orders of magnitude larger than the values observed for transition metals such as Ni, Pd, Pt, W, Fe or Co.\cite{NiPd} Such an enhancement of the $T^2$ resistivity suggests that charge is carried mainly by itinerant $d$ electrons in both AFM and SF phases of MnPtSi.

Remarkably, at temperatures of 280--290~K there are pronounced features in $\rho(T)$ data measured with and without applied magnetic field. Furthermore, $\rho$($T$) is not fully reproducible in the temperature range of $\sim$240--400~K (Fig.~\ref{fig:Fig6}). Similar results were obtained for several specimens with contacts prepared by spot welding and using a silver--filled epoxy glue. 
Although this behaviour may be caused by the presence of grain boundary phase, it is tempting to ascribe it to strain-- and stress--related phenomena in polycrystalline blocks due to structural effects. We note that strongly anisotropic thermal expansions were reported for a number of Mn--based TiNiSi--type compounds, with the largest anomalies in temperature dependencies of the lattice parameters near magnetic phase transitions.\cite{Barcza}

\subsection{Magnetoresistance}
\label{Magnetoresistince}

\begin{figure}
\includegraphics[width=0.45\textwidth,angle=0]{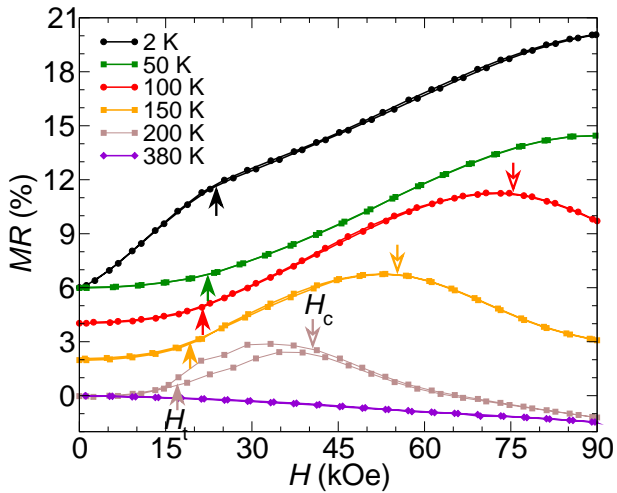}
\caption{\label{fig:Fig7} (Color online) Magnetoresistance ratio as a function of magnetic field at selected temperatures. The MR($H$) curves were shifted along the vertical axis to enable their visual comparison. Arrows indicate phase boundaries $H_{\mathrm{t}}$ and $H_{\mathrm{c}}$  estimated from the magnetization data (Section~\ref{Magnetization}).}
\end{figure}

Fig.~\ref{fig:Fig7} illustrates the magnetoresistance (MR) ratio defined as MR($H$) = ($\rho$($H$) $-$ $\rho$(0))/$\rho$(0), measured with the electrical current transverse to magnetic field. The most striking feature of MR($H$) curves is a prominent broad peak that occurs in the $T$--$H$ region for which the magnetization measurements revealed a SF phase (Section~\ref{Magnetization}). This effect needs to be contrasted with a negative MR expected for magnetic materials due to a decrease of electron--magnon scattering related to damping of magnons by applied magnetic fields\cite{X} and/or caused by a suppression of magnetic superlattice energy gaps. Positive MR was observed for spin--spiral magnets and was attributed to an increase in the number of superzone band gaps during the magnetization process.\cite{Liu-Au2Mn, Takase, Benito, Mackintosh} Therefore, the large positive MR linked to the onset of field--induced SF phase suggests that in MnPtSi the low--$T$ AFM order assumes the form of a non--collinear spin density wave, presumably a helical or cycloidal spin spiral that in applied magnetic field turns into a cone or fan--type spin arrangement in the SF phase. We note that non--collinear spin structures were observed for many Mn--based TiNiSi--type compounds\cite{MnNiGe, MnCoSipierw, MnNiSi, MnIrSi, spinspiral1} and for the parent material MnP\cite{Lifszyc, MnPlast}.

Surprisingly, the shape of MR($H$) curves changes at temperatures between 2~K and 50~K. At $T=2$~K the MR ratio increases quickly with $H$ and peaks near the $H_{\mathrm{t}}$, whereas at higher $T$ the MR ratio is very small in low fields and starts to raise considerably only for $H > H_{\mathrm{t}}$. Further experiments including low--$T$ neutron diffraction measurements as well as magnetotransport study on single crystals are needed to elucidate the origin of the observed change in the MR($H$) curves.

At low temperatures there is no difference between MR($H$) curves measured with increasing and decreasing magnetic field. However, at $T\approx 200$~K a distinct magnetic hysteresis develops near the boundary between the AFM and SF phases, signifying its first--order character. With further increase in temperature MR($H$) curves become irreproducible in the entire investigated field range. Only at temperatures well above the magnetic phase transitions the MR($H$) data shows a consistent, small and steady decrease in resistivity due to the applied magnetic field, as expected when MR is dominated by spin--disorder scattering in a PM state.

\subsection{Hall effect}
\label{Hall}

\begin{figure}
\includegraphics[width=0.45\textwidth,angle=0]{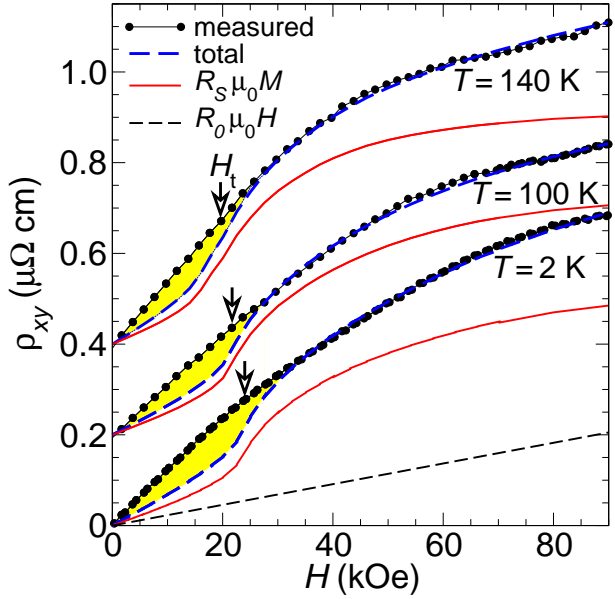}
\caption{\label{fig:Fig8} (Color online) Hall resistivity of MnPtSi as a function of magnetic field at selected temperatures. Dashed blue lines represent fits using Eq.~\ref{eq:Equ3} to the measured data (black dots). The ordinary ($R_{\mathrm{0}}\mu_{\mathrm{0}}H$) and anomalous ($R_{\mathrm{S}}\mu_{\mathrm{0}}M$) contributions are shown as black dashed and red solid lines, respectively. The $\rho_{xy}$($H$) and $R_{\mathrm{S}}\mu_{\mathrm{0}}M$) curves were shifted along the vertical axis to enable their visual comparison. Arrows indicate the AFM--SF phase boundary estimated from the magnetization data (Section~\ref{Magnetization}). Yellow areas visualise the additional contribution attributed to THE.}
\end{figure}

The Hall resistivity $\rho_{xy}$($H$) for MnPtSi presented in Fig.~\ref{fig:Fig8} is strongly non--linear: it shows a positive curvature at low fields and a roughly linear $H$ dependence at high fields. Such a shape of $\rho_{xy}$($H$) curves suggests the importance of AHE. Conventionally, the Hall resistivity of magnetic materials is given by the relation: 
\begin{equation}
\rho_{xy}=R_{\mathrm{0}}\mu_{\mathrm{0}}H+R_{\mathrm{S}}\mu_{\mathrm{0}}M,
\label{eq:Equ3}
\end{equation}
where $R_{\mathrm{0}}$ denotes the ordinary Hall coefficient arising from the Lorentz force acting on the charge carriers, and the second term called anomalous Hall resistivity is a consequence of broken time--reversal symmetry due to a finite magnetization.\cite{Hall1} Attempts to describe the experimental $\rho_{xy}$($H$) curves using Eq.~\ref{eq:Equ3} and the corrected $M$($H$) data gave $R_{\mathrm{S}}\approx~1.6~\times$~10$^{-4}~\mu\Omega$~cm~Oe$^{-1}$ for $T\lesssim100$~K increasing to $R_{\mathrm{S}}\approx~1.9~\times$~10$^{-4}~\mu\Omega$~cm~Oe$^{-1}$ for $T=140$~K, and $R_{\mathrm{0}}\approx~2.3~\times$~10$^{-6}~\mu\Omega$~cm~Oe$^{-1}$. The latter value corresponds to an effective charge carrier density $n_{\mathrm{eff}}\approx$~2.7~$\times$~10$^{22}$~holes/cm$^3$ which is typical for metallic materials with transition elements\cite{Wa1, Wa2, Wa3}.

As shown in Fig.~\ref{fig:Fig8}, a good match between the measured and simulated $\rho_{xy}$($H$) data was obtained for \mbox{$H>H_{\mathrm{t}}$,} but at lower fields the measured Hall resistivities are notably larger than those expected from fits using Eq.~\ref{eq:Equ3} and therefore FM components due to magnetic field--induced spin tilt alone cannot account for the observed AHE in the AFM state. 
The distinct enhancement of AHE suggests the presence of a large fictitious magnetic field (Berry curvature) that generates the so--called topological Hall effect (THE).\cite{THE1, THE2} Recently it was shown that THE can occur in non--collinear antiferromagnets\cite{noncoll, noncoll2, noncoll3}, non--coplanar magnets\cite{noncopl}, and compounds with other non--trivial spin textures\cite{st1, st2, st3, st4} as well as in topological materials hosting Weyl nodes near the Fermi energy\cite{Weyl1}. Since for MnPtSi low--$T$ neutron diffraction measurements revealed a non--collinear spin structure with a small canting angle between magnetic moments of neighbouring Mn species\cite{neutrons}, we attribute the sizeable AHE in the AFM phase to THE due to the non--collinear spin arrangement. Further studies including direction dependent magnetotransport measurements on single crystals and calculation of Berry curvature are needed to elucidate the origin of the AHE/THE in MnPtSi in detail.

\subsection{Electronic band structure calculations}
\label{ValenceBand}

\begin{figure}
\includegraphics[width=0.47\textwidth,angle=0]{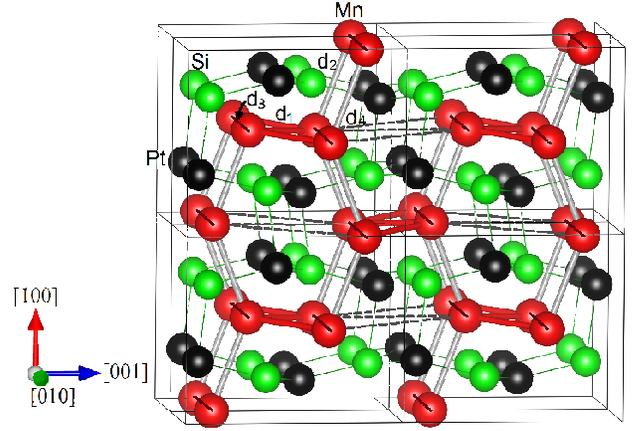}
\caption{\label{fig:Fig9} (Color online) Crystal structure of MnPtSi. Red, black and green balls represent Mn, Pt and Si, respectively. Different couplings between magnetic moments of neighbouring Mn species along $d_{\mathrm{1}}$, $d_{\mathrm{2}}$, $d_{\mathrm{3}}$, and $d_{\mathrm{4}}$ are indicated.}
\end{figure}  

\begin{figure}
\includegraphics[width=0.45\textwidth,angle=0]{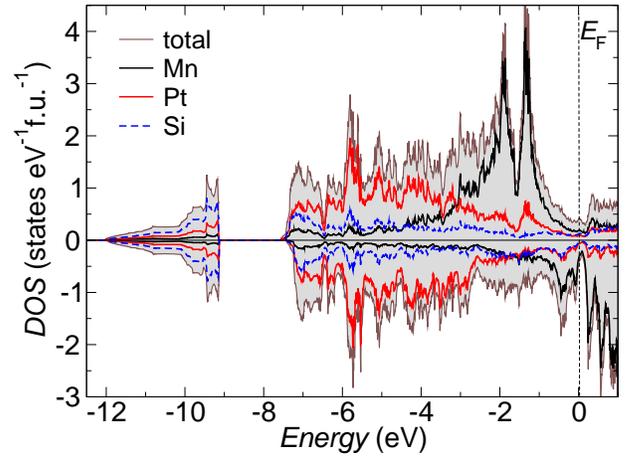}
\caption{\label{fig:Fig10} (Color online) The total and atomic resolved DOSs for MnPtSi calculated assuming the FM state. The majority (minority) spin was plotted upward (downward).}
\end{figure}

First--principles electronic band structure calculations indicate that MnPtSi has a magnetic ground state, in agreement with the experimental findings (Sections~\ref{Magnetization}, \ref{SpecificHeat} and \ref{Resistivity}). Spin--polarized calculations converged into a magnetic solution that has the total energy lower by 833~meV/f.u. than the nonmagnetic state, in line with \onlinecite{PNAS-Rosner}. The computations showed that only Mn species carry considerable magnetic moments. To shed some light on the magnetic structure, various collinear AFM and FIM spin arrangements were studied by means of electronic structure calculations performed on appropriate magnetic supercells. Collinear magnetic structures with moments of neighbouring Mn species coupled ferromagnetically or antiferromagnetically along $d_{\mathrm{1}}$, $d_{\mathrm{2}}$, $d_{\mathrm{3}}$, and $d_{\mathrm{4}}$ (Fig.~\ref{fig:Fig9}) in various combinations were considered. 
Among the simulated non--FM spin arrangements, the lowest total energy was obtained for the spin configuration labelled AFM1, with zig--zag chains along [010] formed by Mn moments coupled ferromagnetically ($d_{\mathrm{1}}$ and $d_{\mathrm{3}}$ in Fig.~\ref{fig:Fig9}), which are coupled antiferromagnetically with each other along [100] ($d_{\mathrm{2}}$ in Fig.~\ref{fig:Fig9}). Nevertheless, the FM solution was found to be energetically more favourable, with the total energy lower by 107~meV/f.u. than that for the AFM1 spin arrangement. Thus, the computational study hints at a more complex magnetic structure, likely non--collinear with a long propagation vector, in line with conclusions drawn from the magnetotransport measurements (Sections~\ref{Magnetoresistince} and \ref{Hall}) and with preliminary neutron diffraction measurements\cite{neutrons}.

The calculated DOSs are similar for all the considered spin arrangements as well as for the PM state simulated using the DLM--CPA method suggesting that the inter--atomic exchange interactions between magnetic moments of Mn have only a minor influence on the overall shape of the resulting DOSs. The computational results indicate that MnPtSi is a metal with a broad valence band (Fig.~\ref{fig:Fig10}). The upper part the valence band is dominated by Mn~3$d$ states, whereas Pt~5$d$ states contribute mostly at the high binding energy part of the valence band, where they hybridize mainly with 3$p$ states of Si. However, there is also a distinct interaction between Mn and the surrounding Pt--Si network reflected in the correspondence between shapes of the partial DOSs of Mn, Pt and Si.

The calculated electron counts for the Mn 3$d$ states equal $\sim$5.3, implying the formal valence of Mn close to 2+. The obtained spin moments are of 3.2~$\mu_{\mathrm{B}}$/Mn, and the orbital contributions are only of $\sim$0.02~$\mu_{\mathrm{B}}$/Mn. Thus, the computational study indicates that Mn in MnPtSi adopts a $J=S=3$/2 configuration, in agreement with magnetization measurements (Section~\ref{Magnetization}). This finding needs to be contrasted with a high--spin state with $J=S=5$/2 expected when $d$ electrons of Mn in a 3$d^5$ configuration keep their atomic--like character and do not participate in bonding, in accord to the Hund's rules. The occurrence of an intermediate spin state indicates that there is a strong interaction between Mn~3$d$ shell and electrons from the surrounding Pt--Si network that competes with the intra--atomic exchange and, as a result, the not--fully polarized spin state arises.

Regardless of the specific magnetic pattern, there is a high contribution of Mn~3$d$ electrons to the calculated DOS($E_{\mathrm{F}}$), as expected based on the temperature dependencies of the electrical resistivity (Section~\ref{Resistivity}). To shed light on the effective mass enhancement in MnPtSi, we estimate bare values of the Sommerfeld coefficient \mbox{$\gamma_b =$ ($\pi^2$/3)$k_{\mathrm{B}}$DOS($E_{\mathrm{F}})$} using the DOS($E_{\mathrm{F}})$ derived from our calculations of 1.17~states/(eV f.u.) for the most energetically favourable among the considered non-FM structures and of 0.57~states/(eV f.u.) for the FM state (Fig.~\ref{fig:Fig10}). The obtained $\gamma_b$ values of \mbox{2.8~mJ/(mol K$^2$)} and \mbox{1.4~mJ/(mol K$^2$)}, respectively, are about two or four times smaller than the experimental Sommerfeld coefficient of \mbox{5.3~mJ/(mol K$^2$)} (Section~\ref{SpecificHeat}), suggesting a moderate mass renormalization due to interactions such as e--ph, electron--magnon and/or electronic correlations.

\section{Conclusions and Summary}
\label{Discussion}

\begin{figure} 
\includegraphics[width=0.45\textwidth,angle=0]{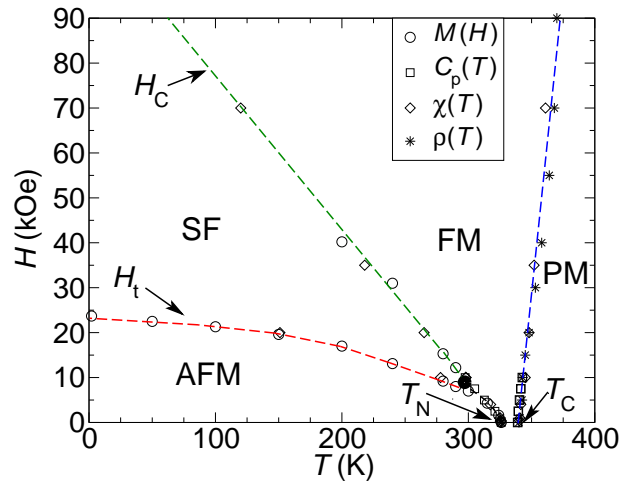}
\caption{\label{fig:Fig11} (Color online) Magnetic phase diagram for MnPtSi. Black dot denotes the critical point at which the SF phase terminates. Dashed lines separating different magnetic phases are guided to the eye.}
\end{figure}

\begin{figure} 
\includegraphics[width=0.45\textwidth,angle=0]{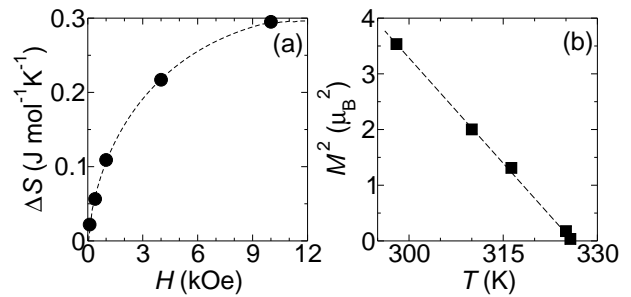}
\caption{\label{fig:Fig12} (a) Magnetocaloric entropy change calculated from the $M$($T$) curves and the \mbox{$dT_{\mathrm{N}}$/$dH$} data using Clausius--Clapeyron relation. (b) $M^2$ vs $T_{\mathrm{N}}$ plot obtained based on the $M$($T$) curves. Dashed lines are guides to the eye.}
\end{figure}

Our study indicates that MnPtSi is an AFM metal with a sizeable $d$--type DOS near the $E_{\mathrm{F}}$. With increasing temperature, it undergoes two magnetic phase transitions, to FM state and subsequently to PM state. The Curie temperature obtained from the modified Arrott analysis, $T_{\mathrm{C}} = 339$~K, agrees with estimates based on the change in slope of the $\rho$($T$) curves at 341~K and the anomaly in $C_p$($T$) at 339~K. Positive curvature of the standard Arrott isotherms near $T_{\mathrm{C}}$ and the $\lambda$--type shape of the specific heat anomaly unanimously point to a second--order character of the FM transition.
In turn, negative slopes of the Arrott isotherms in the AFM phase hint at a first--order nature of the spin--reorientation transition. The latter is evidenced by a peak in $C_p$($T$) at $T_{\mathrm{N}} = 326$~K concurring with the steepest slope in the low--field $M$($T$) curves and a kink in $\rho$($T$) starting at 328~K, with the midpoint at 326~K. Although there is hardly any thermal hysteresis in the $M$($T$) curves near $T_{\mathrm{N}}$ and the peak in the $C_p$($T$) due to the transition resembles a $\lambda$--type anomaly, we note that strain in polycrystalline specimens may influence their critical behaviour by suppressing the latent heat and reducing the thermal hysteresis.\cite{Morrison}

The results of our experimental study are summarized in the $T$--$H$ phase diagram shown in Fig.~\ref{fig:Fig11}.  
Positive slopes of the standard Arrott curves at points corresponding to the phase line $H_{\mathrm{c}}$($T$) (Fig.~\ref{fig:Fig3}) indicate that the SF--FM transition is continuous. In turn, negative curvature of the Arrott lines near the $H_{\mathrm{t}}$($T$) boundary hints at a first--order nature of the AFM--SF transition line. The lack of hysteresis in the $M$($H$) and $\rho$($H$) data (Sections~\ref{Magnetization} and \ref{Magnetoresistince}) suggests that at low $T$ the transition is only weakly first--order. We note that at $T = 2$~K the change in magnetization at the AFM to SF phase boundary, $\Delta M_{\mathrm{t}}$, is only  $\sim$0.6~$\mu_{\mathrm{B}}$ (Fig.~\ref{fig:Fig2}). This, together with a shallow slope of the $H_{\mathrm{t}}$($T$) phase line at low temperatures (\mbox{d$H_{\mathrm{t}}$/d$T \approx$ -25~Oe/K,} Fig.~\ref{fig:Fig11}), is indicative of a small entropy difference between the AFM and SF spin configurations, $\Delta S_{\mathrm{t}}$. Indeed, an estimate based on Clausius--Clapeyron relation \mbox{d$H_{\mathrm{t}}$/d$T$ = - $\Delta S_{\mathrm{t}}$/$\Delta M_{\mathrm{t}}$} gives \mbox{$\Delta S_{\mathrm{t}}\approx$ 8 mJ/(mol K)}.

With increasing temperature the slope of the $H_{\mathrm{t}}$($T$) phase line is getting steeper (Fig.~\ref{fig:Fig11}) and the $\Delta M_{\mathrm{t}}$($T$) is augmenting (Fig.~\ref{fig:Fig2}a), implying that the $\Delta S_{\mathrm{t}}$($T$) is increasing. Remarkably, the magnetocaloric entropy change due to the metamagnetic transition estimated from the $M$($T$) curves (Fig.~\ref{fig:Fig1}) and the \mbox{$dT_{\mathrm{N}}$/$dH$} data (Fig.~\ref{fig:Fig11}) using Clausius--Clapeyron equation peaks near the critical field $H_{\mathrm{cr}}\approx 10$~kOe, as shown in Fig.~\ref{fig:Fig12}a. Therefore, we conclude that the distinct enhancement of the magnetocaloric effect is associated with the critical point at which the SF phase terminates.

The strong magnetic field dependence of $T_{\mathrm{N}}$ (Fig.~\ref{fig:Fig11}) suggests that the AFM to FM transition is accompanied by a magnetoelastic interaction. According to a Landau--type model, the magnetoelastic energy should vary with $M^2$ in a quasi--linear fashion. Therefore, to check for signatures of magnetoelastic coupling in MnPtSi, we estimated $M$($T_{\mathrm{N}}$) values from the $M$($T$) curves measured in different magnetic fields, and we plotted the $M^2$($T_{\mathrm{N}}$) as a function of $T_{\mathrm{N}}$. As shown in Fig.~\ref{fig:Fig12}b, the experimental data supports the presence of a pronounced magnetoelastic interaction. This coupling is believed to be the precursor to first--order magnetoelastic transitions.\cite{Kun}

To summarize, we presented a combined study on magnetic properties and electronic band structure of a novel metallic metamagnet, MnPtSi, in which Mn with the electron configuration of 3$d^5$ is in a not--fully polarized spin state ($J=S=3$/2). Measurements revealed a FM ordering with $T_{\mathrm{C}} = 340$(1)~K followed by a spin--reorientation transition to a non--collinear AFM state at $T_{\mathrm{N}} = 326$(1)~K. The applied magnetic field induces a first--order metamagnetic transition, which is of spin--flop type for $T<300$~K and of spin--flip type for $T$ between $\approx300$~K and $T_{\mathrm{N}}$. Based on the magnetization, specific heat and resistivity data we constructed a magnetic phase diagram with a critical point at the confluence of the AFM, SF and FM phases with $T_{\mathrm{cr}}\approx 300$~K and $H_{\mathrm{cr}}\approx 10$~kOe, near which a peak of the magnetocaloric entropy change is observed. The strong field dependence of $T_{\mathrm{N}}$ and a linear relationship of the $T_{\mathrm{N}}$ with $M^2$ hint at the presence of a pronounced magnetoelastic interaction. In view of these results, it would be of interest to inspect the changes in crystal structure associated with $T$-- and $H$--induced magnetic phase transitions in MnPtSi.

\section{acknowledgments}

The authors are grateful to the Dresden High Magnetic Field Laboratory for support with the high magnetic field measurements. M.G. would like to acknowledge financial support from the Max Planck Society through research fellowships.


\begin{thebibliography}{}
\bibitem{note}
The demagnetization correction $H_{\mathrm{int}}$~=~$H$~-~4$\pi~N~M_{\mathrm{vol}}$ for the factor $N$~=~0.24 has been applied where $N$ was determined from the shape of the pollycrystalline piece. Similar value $N\approx$~0.2 was obtained from the inverse slope of $M$($H \rightarrow$~0) at $T=2$~K.
\end{thebibliography}
\end{document}